# Measurement, Analysis, and Insight of NFTs Transaction Networks

Prakhyat Khati
Computer Science
University of Saskatchewan
Saskatoon, Canada
khati.prakhyat@usask.com

## ABSTRACT

Non-fungible tokens (NFTs) are unique digital items with blockchain-managed ownership. Ethereum blockchain-based smart contract created the environment for NFTs (ERC-721) to reach its one of the most important future application domains. Non-fungible tokens got more attention when the market saw record-breaking sales in 2021. Virtually anything of value can be traced and traded on the blockchain network by minting them as NFTs. NFTs provide the users with a decentralized proof of ownership representation, as every transaction and trade of NFTs gets recorded in the Ethereum network blocks. The value of NFTs is derived from their being "non-fungible," meaning that the token cannot be replaced with an identical token (giving it inherent scarcity). In this paper, we study the growth rate and evolutionary nature of the NFT network and try to understand the NFT ecosystem. We explore the evolving nature of the NFT interaction network from a temporal graph perspective. We study the growth rate and observer the semantics of the network. Here on the observer network, we will run two graph algorithms on the dataset. Lastly, observe and forecast the survival of NFTs bubble by applying the Logarithmic periodic power law (LPPL) model to the time series data on one of the most famous NFT collections," CryptoPunks" (predicting price increase), which has seen sales of around $23.7million around mid of 2021.

## CCS CONCEPTS

• Mathematics of Computing→Graph algorithm: Exploratory data analysis. Time series analysis.

## 1 INTRODUCTION

Blockchain is a distributed database that is shared among the nodes of a computer network. It allows digital information to be recorded and distributed. Blockchain and cryptocurrency have been tightly coupled since the birth of blockchain technology with Bitcoin. Ethereum is an upgraded blockchain in terms of scalability and functionality compared to Bitcoin[1]. Ethereum paved a new way by introducing an automation layer on top of a permissionless blockchain fabric, using complex smart contracts executed by a decentralized network. This led to blockchain development into its current avatar, where decentralized applications are written on the framework. Ethereum blockchain is a transaction-based state machine, where the stare is made up of accounts. The transfer of values and information between accounts causes transitions in the global state of Ethereum, which are recorded in the blockchain[2]. Understanding how the distributed blockchain works is quite complex. There are different entities involved, such as miners who create blocks by verifying the transaction and adding the transaction to the distributed ledger by solving the consensus algorithm. In addition to the native unit of value ether, the Ethereum blockchain allows the use of the tokens-an abstraction of digital assets, with the help of suitable data structures and methods implemented through a smart contract. Appropriate smart contracts can be used for implementing various types of tokens. Smart contracts are the heart and soul of cryptocurrencies and other tokens. NFTs are governed and stored by smart contracts, and these smart contracts will lay the foundation and terms of sale[3]. In the Ethereum blockchain, ERC-20 [4]standard applies for a fungible token,[5] like the ETH [6](ether). If an ERC-721[7] standard is applied, then we are dealing with NFTs.

Non-fungible tokens are unique digital assets on a public blockchain[8]. Unlike the fungible token, NFTs introduced standardization, interoperability, tradability, liquidity, immutability, provable scarcity, and programmability[9]. All these disruptive elements made NFTs personal property, not contract or pure intellectual property license.[10]. NFT applications are now a hot research topic, but till today context NFTs are linked to a digital context. Digital arts are one of the first applications of NFTs to showcase the item owned and the ability to trade those items[11]. The year 2021 was NFTs year; there were more than a million transactions on the Ethereum blockchain network each day just for NFTs only. The trading currency used was either ETH or WETH.

If we change that into fiat currency such as dollars, then all those trade transactions were worth millions of dollars each day. In April 2022 the overall market capitalization of the top NFTs collection is around $13,492,390,000[12]. The instant growth naturally led people to question whether the current NFT market is a bubble[13] or not. To answer this question, one of my research questions aim to identify and predict the bubble of one of the popular NFT "CryptoPunks" market from its empirical data. We used Logarithmic Periodic Power Law (LPPL) model [14], [15].

Also, Here the NFT transaction network in the public blockchain also brings a fascinating ecosystem of humans (Traders/ Users / Minters) and autonomous agents(contracts). This network is neither like an online social network, where players are all human users but also not a complete core financial network, where all the interactions are transferred or value or assets. In the NFT network, unlike the financial network, the assets could be anything unique that has some signifying values. The NFTs network is close to the Internet or Web, where users and programs are allowed to interact with one another, following predefined rules of engagement. There has been little work done to understand the evolution and temporal properties of the network of NFT transactions, observe and identify the trust or influential users(wallets)[16], and gain some in site into NFT-related communities in terms of their interaction graph and associated properties. In our NFTs transaction graphs, the nodes are users or traders, and edges are the transaction between these nodes. These transactions could be done by a smart contract or a user. We aim to address their two main research questions.

**(1)** How do NFT transaction networks evolve over time, what are the network properties, and identify nodes that influence the market over different NFT collections and visualize them.

**(2)** Identify and predict the "CryptoPunks" collection are generally a medium bubble (predicting price increase) using the time series data.

We systematically investigated 7 NFT collection (ERC-721) transaction ecosystems from 2017 to 2021 via graph analysis. The paper consists of 6 sections, including this introduction. Section 2 will provide some background on Ethereum and ERC-721 related works to bubble prediction and graph analysis. Section 3 consists of datasets extraction, preprocessing, and experimental setup. Section 4 show the methodology involved; Section 5 will discuss the limitation and challenges; finally, Section 5 will conclude this study while mentioning future works.

## 2 RELATED WORKS

Graph analysis of Blockchain networks. Several works explored cryptocurrency networks based on graph theory and network analysis. Among them, BitCoinView[17] was developed to visualize the flow of the bitcoin transaction graph. This can be used to understand what essential things are needed and to be considered to visualize a transaction. Similarly, numerous algorithms have been developed to discover the relationship between smart contracts and tokens[18] and understand the ecosystem of Ethereum using graph analysis.

Very recently, S. Casale-Brunet[19] analyzed the NFTs communities; the author selected the top 8 NFT collection projects and claimed that the NFT network graph follows power-law in their degree distribution; they also compared the topological values with social network and web networks, ERC-20 token networks. Our approach is quite similar, but we have added Temporal analysis of the NFT network and predicted the network bubble. Similarly, to understand a node's influence in the blockchain network. Maeng, Soohoon Essaid, and Meryam [20]have worked on the visualization of Ethereum p2p network topology and analyzed the attributes of the network such as node degree, path length, diameter, and clustering coefficient. They then explore the node properties and provide analytical results regarding the relationship between nodes, heavily connected nodes, node geo-distribution, security issues, and possible attacks over the influential nodes.

Network properties of transaction graphs. Bernhard Haslhofer, Roman karl and Erwin Filtz[21]studied the large-scale network properties of Bitcoin transaction graphs. Similar work was done by Ron D and Shamir A[22]. D'angelo studies only the transaction of the Ethereum network[23]. There has been numerous network analysis of the ERC-20 token. Somin et al[24] investigated the entire address graph spanned by ERC20 token trade-in the Ethereum blockchain and studied the social signal in the Ethereum ERC20 token trading network.[25]. Victor and Luders recently measured Ethereum based ERC-20 token network. [26]. Much less work has been done in the NFT ERC-721 trading network. S. Casale Brunet, P. Ribeca, P. Doyle, and M/. Mattavelli was among the first to propose a systematic analysis of the transaction network. [11].

All the transaction carried on the blockchain is publicly available; because of this, empirical study of NFT markets and ideas for implementation in different sectors can be seen. Recently Nadini, et al is the first empirical study on NFTs. They have analyzed a massive 6.1 million NFT trade transaction between 2017 and 2021[27]. They also have mentioned ways to detect communities and analysis them. Ante l[28]studied the relationship between Bitcoin and Ethereum and NFT sales.

Bubble prediction on Cryptocurrencies. Bubble prediction on big cryptocurrencies, such as Bitcoin, Ethereum, and Solana. Basically, having high market capital has been a popular topic. One of the early from 2014, Macdonell A[15] is the first paper to study this topic of bubble prediction. Here the author used the LPPL model as well. Since then, there have been numerous research on bubble prediction. Buanchetti M and ItoK, shibanoK [29], [30] studied different ways of using the LPPL model in other cryptocurrencies. In most of the papers, the commonly used model is the LPPL model. That is the reason behind the chosen LPPL model for the NFT bubble prediction. Some of the papers have tried adding new terms to the model and tried fusing models to observe a new finding. [31]Though the commonly used model is the LPPL model. Some of the research attempted to use the machine learning approach [32] to predict the price of the cryptocurrency price bubble[33].

There is not much work done on bubble prediction of the NFT bubble. The NFTs marketplace, such as Open Sea, shows a strong growing market, but the evaluation of the NFTs collection market is lacking. Very recently and almost concurrent to ours [34] analyzed the top 5 NFTs and predicted the bubble for each NFTs collection as per their size. Our approach is quite similar, but we are using the LPPL model only for a single NFTs collection

## 3 DATASETS AND EXPERIMENT SETUP

Different blockchain supports NFT trading, but we focus only on NFTs that are traded in the Ethereum blockchain network. Ethereum, as mentioned previously, is a public distributed ledger. All the transactions done are available publicly. But these transactions have a lot of metadata on them. We explored all the Datasets using the Google cloud big-query repository[35]. Google Bigquery is a data warehouse that can handle large-scale data and make it easy to access using SQL syntax. The data set consisted of 7 different NFT collection tables. We searched for information on the blockchain by directly accessing records(blocks) using a unique identifier (e.g., block ID, Transaction, wallets, and contract address). The file was stored in a comma-separated (CSV) file. The files include information such as truncation hash, address of the NFT smart contract, and the address of the wallet(s) or smart contract (both buyer and seller).

### 3.1 Extending the data set.

Alongside the dataset obtained from the big query, we also used two other datasets collection. One of the NFT collection data sets was used from a scientific report[27] done under the NFT revolution. This data is used to observe the transaction and visualize the NFT network. This is a huge data set, and it contains 6.5 million transactions for around 4.6 million NFTs. The NFTs are part of 4600 collections and 6 categories. Each transaction has a date and the price in crypto and USD; each transaction describes which NFT was sold. This is really huge data, so to dig deeper, we observed visualization on a subset of the dataset. The subset data set consists of transaction done in a single day, 2021-01-01, as even in a single day, there is quite a lot of transaction. This dataset was available in the .dump file only and needed to be imported using the neo4j desktop application as a database. So, a Graph schema model needs to be designed, and every column needs to be allocated to the respected nodes, as shown in Fig 1. Neo4j Data importer is used to import the data and create a graph model. Neo4j desktop version or Neo4j Aura, the web version, can handle such large data, and to query the dataset, Cypher is used. Neo4j handles the relational database efficiently and is quite a popular tool for bigdata analysis. Fig 2 is the updated version of the same graph schema that divides the Trader node into Buyer and Seller nodes.

Similarly, to calculate the NFT-CryptoPunks bubble prediction, the time series data is used. This data is retrieved from non-fungible.com[36]. The CSV file consists of two columns; one is the date, and the other one is the average weekly price. The time-series data is calculated for CryptoPunks only, and it is starting from mid-2018 to 2021. The price data is associated with the collection of the NFT rather than each NFT.

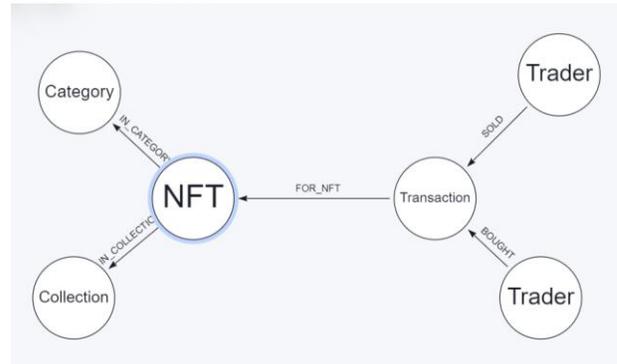

Fig 1: Graph Schema Model of NFTs

We consider the price of the NFT collection together rather than the individual NFTs. In other words, we assumed the NFTs are homogeneous for simplification. We will be using the LPPL model on the same data set.\

## 4 METHODOLOGY

Our whole analysis is based on the NFTs network and its network components. So first, we need to analyze the data itself to under more about the distributed nature of the NFT network. In the fig3 below, we can see that the overwhelming number of NFT owners each only own a small number of tokens.

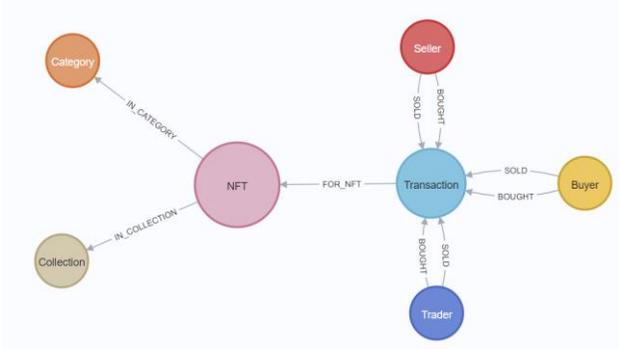

**Fig 2: Modified Graph Schema Model of NFTs**

There are very few addresses that own hundreds or even thousands of tokens. Their addresses are the collection address and are common to all the NFTs in the collection. Fig 4 is charted on a logarithmic scale for easy interpretation. This reflects the distribution of the number of tokens per address seems to follow a Zipfian distribution, as indicated in fig 4. Form the figure, any address which owns thousands of tokens is either purchasing those tokens automatically (using smart contracts) or is financing the collection in which they own the tokens. First, we analyze the ownership trends addresses that do not own many tokens.

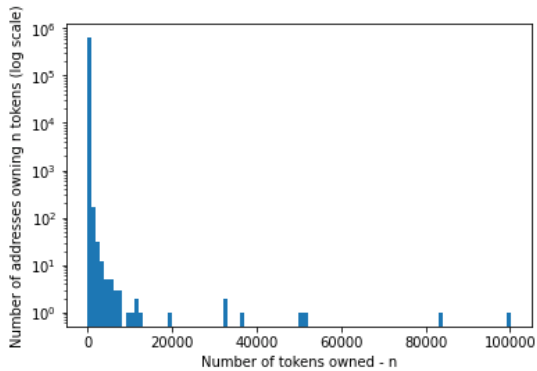

**Fig3: Number of addresses owned.**

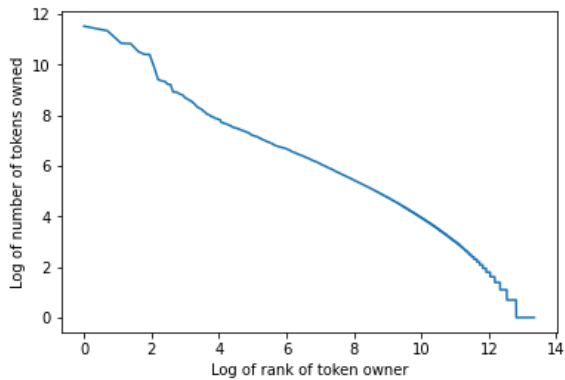

**Fig4: Number of tokens owned.**

This finding will help us to estimate trends in NFT ownership. So here, some of the address contains a lot of tokens, so we select a cutoff value to be 1500. Fig5 shows that the decentralized NFT market is indeed decentralized. Most of the NFTs are digital arts now, and these digital arts get released in the collection. So, all the price information, owner information, and other metadata would be handled by a single contract account for this collection. Few of the NFTs collections are more than just a digital art

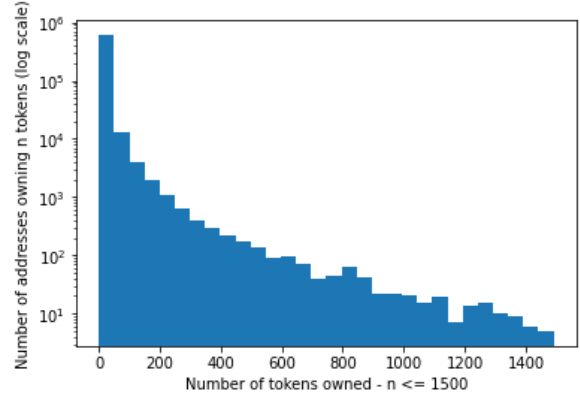

**Fig5: Number of tokens owned (the cutoff point)**

For example, NFT like "ENS" (Ethereum Name Service)[37], which behaved like a DNS, and this naming service can be used to a greater extent. NFT collection is unique, and each collection has different significance and values.

### 4.1 NFTs Transaction Network

We created the interaction network to perform the transaction analysis on the NFT network, based on the extracted table tokennet table from dataset1 observed from the bigquery. This table is built based on the token transfers from one address to another address. Their address can be of buyer or seller. Some of the addresses 0x000 denote mint and burn addresses. Whenever an NFT is removed from the blockchain ledger, then the destination address will be the burn address, and whenever an NFT is minted or created to the collection, its address will be 0x000 initially before an owner is assigned. Like our network, we structure the transaction graph model in a multidirector weighted graph MDG(V, E) with a set of nodes V and edges E. Each node represents a single wallet address; it can be considered a user address. Each directed edge E represents a single token transaction between the nodes V. For each transaction, information is defined and stored as edge parameters. For example, the transaction_hash_id cost paid by the wallet, Date_time of the Transaction. Here we used Python" igraph" package to extract the relevant information from the data table.

The sale of NFT skyrocketed in 2021; fig6 shows the average amount of sales over time. We can observe the table1 which shows the year's increase in the number of transactions. In 2018 there were only 450 thousand transactions, whereas in 2021, there we almost three and a half million transactions. If we look at table 1, we can see a boom in NFT world as in the first quarter only; the total volume has been rising to nearly 760 million USD. Here the indegree and outdegree of each node is too large to visualize; the fig7: shows the top 10 accounts that have high degree value. This high degree value means that these addresses are active in trading the NFTs. The highest number of NFT is, as shown in the fig7 is held under an address starting from 0x0000. This means that any user does not own this address; these addresses are associated with NFTs when they are the first minted.

We selected the major 7 NFT projects [37] of all time as per the number of sales over the period and calculated the semantics of the combined network by performing graph analysis on the transaction data. We observed that there was a total of 33678 unique addresses, which represent nodes of the graph, similarly 425246 number of transactions between those addresses.

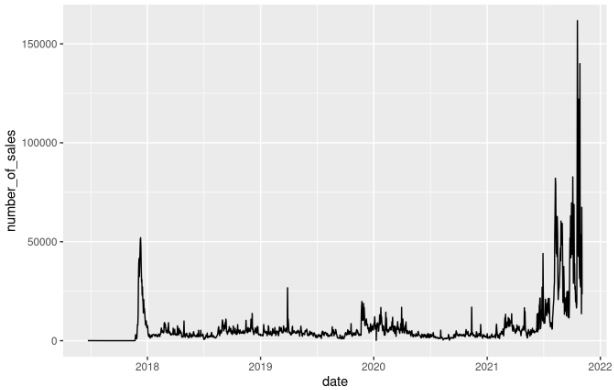

**Fig 6: NFT daily number of sales.**

The constructed network is a directed graph. We calculated the graph network reciprocity, associativity, connected component and k-core properties of the network. The reciprocity of a directed graph gets the likelihood of nodes in a directed network being mutually linked.

$$r = |(u,v) \in G | (v,u) \in G | / |(u,v) \in G|.$$

Here the ratio is the number of edges pointing in both directions to the total number of edges in the graph. This shows the trade transaction between the user nodes. Similarly, in the above equation. The reciprocity of a single node u is defined. It is the ratio of the number of edges in both directions to the total number of edges attached to node you. G denoted the directed graph. The reciprocity of the graph was observed to be 0.**068361305**. Similarly, we found out among the nodes, 18279 were strong nodes and 72922 strong, connected edges between the nodes, and we observed that the max weak nodes were 95 nodes only; this shows how well connected all the nodes are. We observe that the max weak connected components edges counts were 104064. Table 2 gives all other graph parameters that are observed from the NFT transaction graph. Table 2 shows strong, connected nodes and edges count. The assortativity is calculated in an indirection graph.

| | year | transactions | totalVolume | averagePrice |
|---|---|---|---|---|
| 0 | 2017 | 253100 | 1.829279e+07 | 72.274933 |
| 1 | 2018 | 449373 | 1.667865e+07 | 37.118441 |
| 2 | 2019 | 746489 | 1.867730e+07 | 25.086969 |
| 3 | 2020 | 1245954 | 7.762350e+07 | 62.551672 |
| 4 | 2021 | 3376111 | 7.559576e+08 | 223.996997 |

**Table 1: Total Transaction of NFTs.**

Here we first combined the 7 different CSV into a single collection and assigned each address a unique id. giant () method was used to get the strongly connected components, a similar approach for the weak components. Then we calculated the core from the large weakly connected components and used the k_core method to obtain the number of strong and weak connected component vertex and edge count.

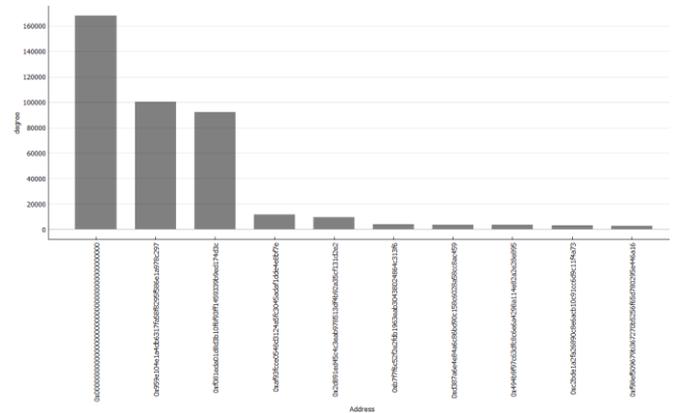

**Fig7: Top 10 account address with high degree value.**

| Reciprocity | 0. 068361305 |
|---|---|
| Assortativity | 0.092805781 |
| Strong nodes | 18279 |
| Max scc node | 15151 |
| Max scc edge | 72922 |
| Weak nodes | 95 |
| Max wcc node | 33455 |
| Max wcc edge | 104064 |
| Wcc -nodes | 148 |
| Wcc -edges | 2239 |
| Scc -nodes | 148 |
| Scc -edges | 2239 |

**Table 2: Transaction network semantics.**

The below figure shows the ALL NFT network component as shown here the buyer, and the seller are created from the trader, and the transaction is considered a component for better visualization famous artist are sold at high prices. In a similar manner, the collection of NFT and who created the NFT determines the worth of the NFT

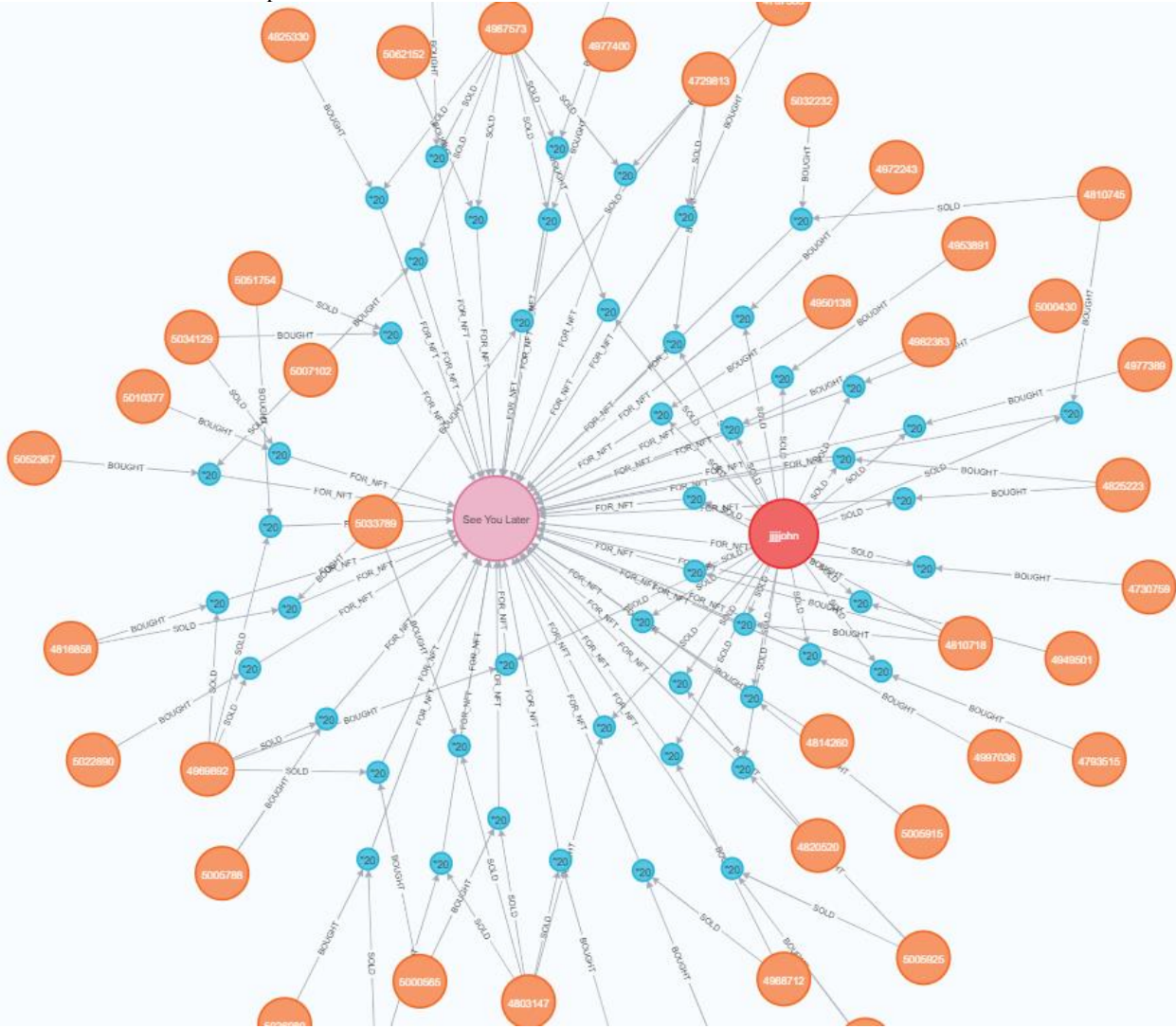

**Fig 8: NFT trading transaction Graph.**

As shown in fig2. Let's us take an example to visualize the transaction of a Single NFT transaction over the period. As we can see in the below figure. The pink color is an NFT called "**See You later**" and the red is the create of the NFT and then all the orange nodes are the buyer and seller from the time of creating the NFT. The blue color represents the transaction and the kind of transaction that happened. For example, it shows which wallet sold the NFT and which wallet bought the NFT from that transaction.

We can see the graph schema contains the collection components. Most of the popular NFT belong to a collection. We can see the example in real-world where most of the famous paintings of a Collection. There is not any price limit as there is no condition on how much the price can be set.

| | collection | averagePrice | numberOfNfts |
|---|---|---|---|
| 0 | Saturdaynightlive | 360856.408928 | 1 |
| 1 | Appresidential | 181832.879943 | 1 |
| 2 | Trippderrickbarnesxflipkick | 118838.578811 | 1 |
| 3 | Fairumnft | 99849.750000 | 1 |
| 4 | Chainsaw | 96546.024285 | 12 |

**Table3: Top 5 NFTs collection as per price.**

| | username | bought | sold |
|---|---|---|---|
| 0 | 0x76481caa104b5f6bccb540dae4cefaf1c398ebea | 130231 | 192586 |
| 1 | 0x327305a797d92a39cee1a225d7e2a1cc42b1a8fa | 0 | 149142 |
| 2 | 0x4FabDA | 28 | 49338 |
| 3 | 0xfc624f8f58db41bdb95aedee1de3c1cf047105f1 | 1976 | 43571 |
| 4 | StrongHands | 43384 | 437 |

**Table 4: Top 5 NFTs trading address.**

Similarly, to get furthermore InSite on to influence nodes in the NFT network, we used Article Rank algorithm, which is a variant of the PageRank algorithm. It measures the transitive influence of nodes.

Neo4j provides us with built in methods to calculate the ranking. Here the Article rank lowers the influence of low degree nodes by lowering the score being sent to their neighbor in each interaction.

$$ArticleRank_i(v) = (1-d) + d \sum_{w \in N_{in}(v)} \frac{ArticleRank_{i-1}(w)}{|N_{out}(w)| + \overline{N_{out}}}$$

Where, $N_{in}(v)$ are the incoming neighbors and $N_{out}(v)$ are the outgoing neighbor of node v. Similarly, d is a damping factor in [0,1] usually set to 0.85 and $\overline{N_{out}}$ us the average outdegree.

| username | score |
|---|---|
| YellowHeartFactories | 81.98406867254229 |
| Pranksy | 35.33914246905004 |
| 0xfc624f8f58db41bdb95aedee1de3c1cf047105f1 | 27.586534417053834 |
| 0x4FabDA | 23.406713333127424 |
| ethernitychain | 17.147044455053717 |

**Table 5: Top five high influencer NFTs**

| boughtCount | soldCount | boughtVolume | soldVolume |
|---|---|---|---|
| 3 | 5423 | 5176.7485000000015 | 709673.2349614704 |
| 3044 | 20198 | 3931822.8205273137 | 9985688.863616494 |
| 1976 | 43571 | 129906.86449180117 | 355170.4146149685 |
| 28 | 49338 | 68200.18175847222 | 542552.2707045466 |
| 0 | 1055 | null | 628666.0000544671 |

**Table 6: Top five high influencer trading data.**

The Table 5 & 6 show the most influencer NFT collectors of the network and their trading value. The sold and bought volume is in USD. This shows there is signifying amount of transaction influence by these nodes. In the Fig 9 we can observe a community of such influence nodes distinct using a different color. We can observe a denser community at the center and multiple such high influence nodes and multiple small communities being formed around them and these nodes are densely connected signifying high transaction between the nodes.

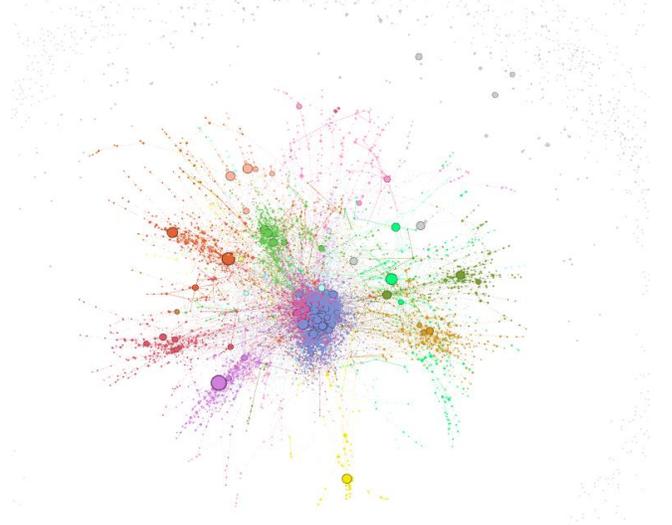

**Fig 9: Community detection of Influencer nodes.**

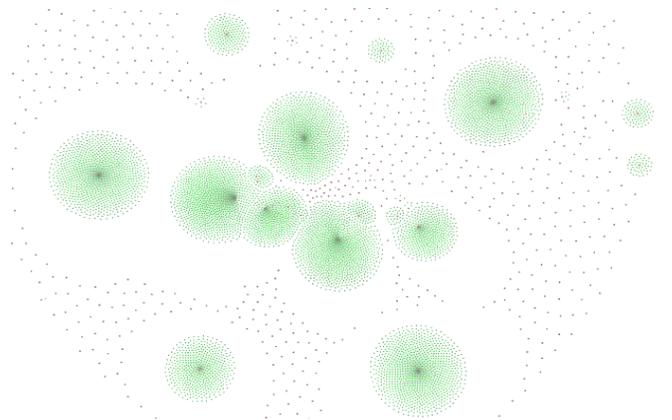

**Fig 10: NFT-Collection visualization**

## 4.2 NFTs Bubble Prediction

In recent years, the LPPL model[38] has been studied in different research for bubble detection, here, the model uses time-series price data. This model assumes the presence of two types of agents in the market: a group of traders with rational expectations and the second group of so-called "noise" traders, that is, irrational agents with herding behavior. The algorithm approximates the log price of the data at a given period t. as

$$f(x) = A - B(T-x)^m (1 + C \cos(\omega \ln(T-x) + \phi)),$$

where A, B and C are the linear parameters and other are the nonlinear parameters Tc, m, w, φ. In the above equation, Tc denotes the singularity at which the previous bubble ends and transitions to another set. A denotes the log price, B denotes the magnitude of the power law acceleration and ranges from B>0 for the increasing price to B<0 for decreasing price. The LPPL model derives these positive and negative bubble indicators to all t2, thereby visualizing its own predictions. Similarly, C denotes the magnitude of the log-periodic oscillations, where if m=1, C=0, then LPPL reduces to an exponential fit of the price time series. If m<1, then LPPL grows super-exponentially until the critical time T. If C, w>0, then LppL exhibits oscillation that becomes progressively more frequent as x approaches the critical time. Here for the model, we need to calibrate the model. Model calibration can be defined as finding a unique set of model parameters that provide a good description of the system behavior and can be achieved by confronting model predictions with actual measurements performed on the system. For our test the model is calibrated with the Ordinary Least square method, providing the estimations of all parameters mention in the above equation. Similarly, the equation also contains non-linear parameters. we set the following calibration for a given time [t1, t2]. We then make bubble predictions at each period of the data, by letting this calibration iterate for the shrinking time window [t1, t2].

$$\max\left\{t_2 - 60, \frac{t_2 - 0.5}{t_2 - t_1}\right\} < t_c < \min\left\{t_2 + 252, \frac{t_2 + 0.5}{t_2 - t_1}\right\}$$
$$0 < m < 1,\ 2 < \omega < 15.$$

According to the daily data, [t1, t2] is in daily units. t2 denotes a fictitious today corresponding to t; t1 denotes an earlier day. For a given t2, the iterative calibration sets the initial range of the time window as 120 days and the shrinking Interval of t1 as 5 days. That is, we need to estimate parameters 24 times for each t2 (e.g., [1, 120], [5, 120], …, [115, 120]). This process emphasizes the prediction and here we use only historical data as input. The outcome of t2 depends only on data from t2 to the last 120 days.

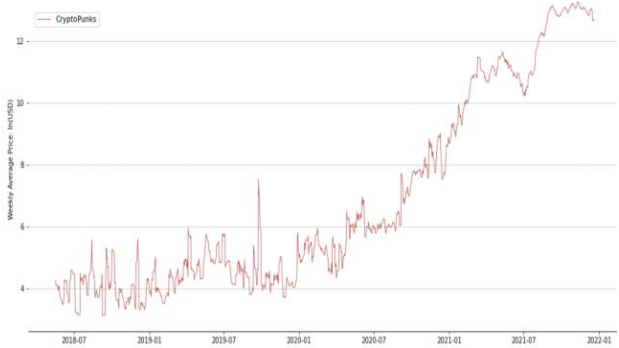

**Fig 11: Time-Series price data used in our analysis**

## 5 DISCUSSION AND CONCLUSIONS

In this work, we investigate the NFT transaction graphs and analyze the different global network properties e.g., reciprocity, assortativity, core decomposition, and clustering coefficient, we realize how their connections change. We discussed what these network properties resemble in such a network. We observed the log rank of token owners and found it to be following Zipf's law which is generally understood to be a power-law distribution with integer values. Many real graphs for social media and the internet show existence of (several)high degree nodes. For example, several social networks have influencer nodes, and in web networks, they are popular and high-ranked websites. These all kinds of networks have been confirmed to follow power laws in their degree distribution. We observed high influencer nodes in the network and community being formed around them, indicating single hub dominating the network. The single hub is the NFT collection owner who are frequently trading. In face in [5].it has been shown that a single hub frequently dominates individual networks for ERC-20 tokens. Article ranking was used to identify such influencer nodes and Gephi was used to plot the community formed around such nodes. This led to the conclusion that the structure of NFT network is qualitatively very similar to the one measured for interaction in social network

In the fig 12 below, overall, the LPPL model seems to capture the the trend of both the positive and negative bubble for CryptoPunks collection. In the result, although the price rose further after bubbleindicator(pos) reached its highest in late August 2020. The model is successful in predicting the direction of price change but around October 2020 to March 2021 here the *bubbleindicator(pos)* failed to predict the continuous price increase Similarly, from the fig 12, near December 20,2021 the bubble indicators are signaling negative value ~ 0.3. This implies that NFT collection are in general, a small bubble (predicting price increase)

This study also shed quantitative light on a NFT market that might otherwise be prone to hype and misleading information.

**Future work**

Following our characterization of the NFT network, there is ample opportunity for future work. Study the nature of NFT collection, identify what distinguishes the NFTs collection and what adds on value to those collections. Identifying entropy of ownership could be one of many ways to capture utility of NFT collection. Similar Further research can be used to refine our knowledge and understand the relation between blockchain platforms and NFT collections specific to those platforms. Study the temporal nature of network community formation of NFTs and predict the nature of the communities Similarly bubble prediction of the whole NFT collection can be done. Also, in the current time-series we assumed the CryptoPunks collection to be homogeneous nature, but all these NFTs are heterogeneous. This is one of the main characteristics of

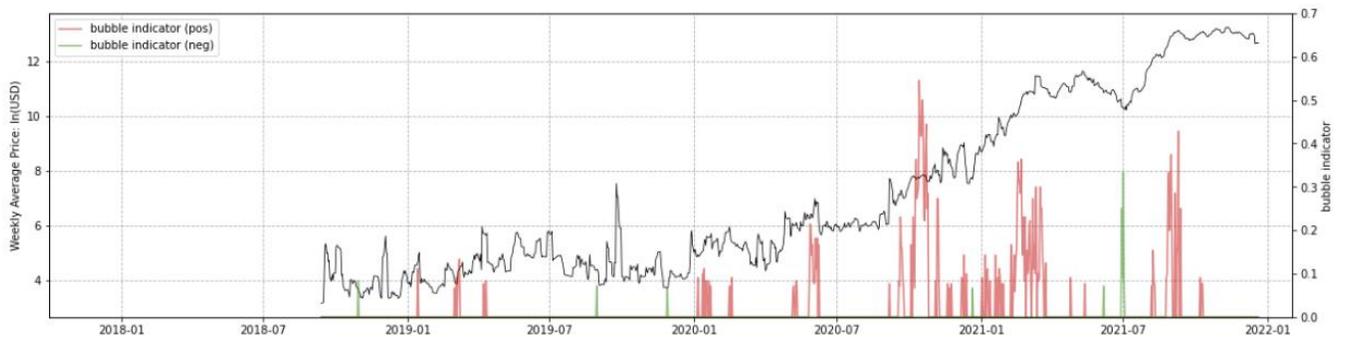

**Fig12: Bubble prediction of CrypotPunks NFTs collection.**

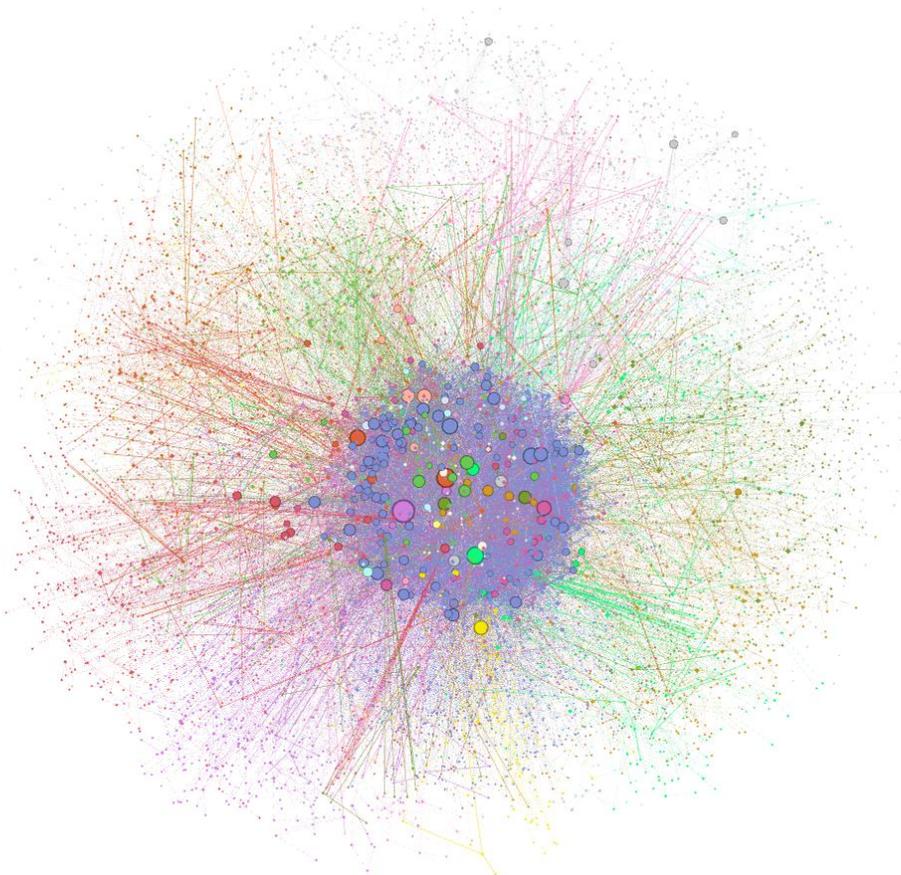

**Fig 13: Community Detection Visualization of NFT Network**